\documentclass[prb,twocolumn,showpacs,floatfix]{revtex4}
\usepackage{amsmath,amsfonts,amsthm,graphicx}
\usepackage{graphicx}
\usepackage{graphics}
\usepackage{dcolumn}
\usepackage{bm}

\newcommand{\beq}{\begin{equation}}
\newcommand{\eeq}{\end{equation}}
\newcommand{\beqnar}{\begin{eqnarray}}
\newcommand{\eeqnar}{\end{eqnarray}}
\newcommand{\bfig}{\begin{figure}}
\newcommand{\efig}{\end{figure}}

\usepackage{color}
\usepackage[dvipsnames]{xcolor} 

\begin{document}
\title{Effect of asymmetric Fermi velocity on trigonally warped spectrum of bilayer graphene}
\author{Fatemeh Adinehvand, Hosein Cheraghchi}
\email{cheraghchi@du.ac.ir} \affiliation{School of Physics,
Damghan University, 36716-41167, Damghan, Iran}
\date{\today}
\newbox\absbox
\begin{abstract}
We derive an effective Hamiltonian at low energies for bilayer graphene when Fermi velocity manufactured on each layer is different of the velocity measured in pristine graphene. Based on the effective Hamiltonian, we investigate the influence of Fermi velocity asymmetry on the band structure of trigonally warped bilayer graphene in the presence of interlayer applied bias. In this case, the Fermi line at low energies is still preserved its threefold rotational symmetry appearing as the three pockets. Furthermore, the interlayer asymmetry in Fermi velocities leads to an indirect band gap which its value is tunable by the velocity ratio of the top to bottom layer. It is also found that one of the origins for emerging the electron-hole asymmetry in the band structure, is the velocity asymmetry which is large around the trigonal pockets.
\end{abstract}

 \maketitle

\section{Introduction}
One of the great interest to study bilayer graphene comes back to its potential application in nano-electronics, thanks to opening a tunable gap by means of an applied electric field\cite{falko2006,mccann2006,ohta2006,castro2007,oostinga2008,zhang2009,Kruczynski2010}. Low energy band structure of bilayer graphene undergoes an asymmetric shaped named as trigonal warping (TrW) for energies lower than the Lifshitz energy\cite{falko2006,Kruczynski2010,das2008, mccann2007, Koshino2006, Cserti2007}. TrW is originated from the skew interlayer hopping integral between $A_d-B_u$ atoms in a Bernal stacking bilayer graphene. Although TrW is effective for only very low energies (of the order of 1 $meV$) in the dispersion, it has a significant impact on the minimal conductivity such that conductivity increases to six times as large as that for single layer graphene\cite{Cserti2007,Moghaddam2009}. In the absence of inter-valley scattering, TrW causes to suppress weak localization\cite{Falko2007}. Moreover, TrW affects magnetotransport and scaling properties of a bilayer graphene in the Corbino geometry\cite{Rycerz2016}. In a junction containing of two trigonally warped bilayer graphene in which two biases are applied with opposite polarities, two topological confined states emerge inside the gap\cite{Falko2015}.

The electronic properties of graphene can be modified by many scattering factors such as structural deformation\cite{Verhagen2015}, doping\cite{Attaccalite2009}, dielectric screening\cite{Siegel2011,Jang2008} and electron-electron interaction\cite{Trevisanutto2008,Elias2011,Bostwick2007} which give rise to a change in the Fermi velocity of charge carriers in each graphene layer. Some of structural factors can be enumerated as strain\cite{Verhagen2015}, modification in curvatures of graphene sheets\cite{Du2008} or generating graphene superlattices by means of a periodic potentials\cite{Park2009,Yan2013,Park2008}, lattice incommensurability with the substrate\cite{Ortix2012}, a twist in bilayer graphene\cite{Yin2015} or folded graphene\cite{Ni2008}. So a modification in Fermi velocity of carriers induced by strain and curvatures such as nanoscale ripples or envirnmental factors is an inevitable event during an experiment especially for a bilayer graphene grown on a substrate in which layer symmetry breaks\cite{Falko2013}.

A change in Fermi velocity of Dirac fermions in monolayer graphene only leads to a renormalized Dirac cones\cite{Raoux2010} while an interlayer velocity asymmetry applied on bilayer graphene induces an indirect band gap which is controllable by the velocity ratio of the upper to lower layers\cite{cheraghchi2014}. Let us mention that our last study on band structure modification induced by velocity asymmetry was focused on high energy excitations with $4 \times 4$ full Hamiltonian. However, the influence of velocity asymmetry on TrW of bilayer graphene at low energies around Lifshitz energy has not been investigated.
\begin{figure}
\centering
\includegraphics[width=8cm]{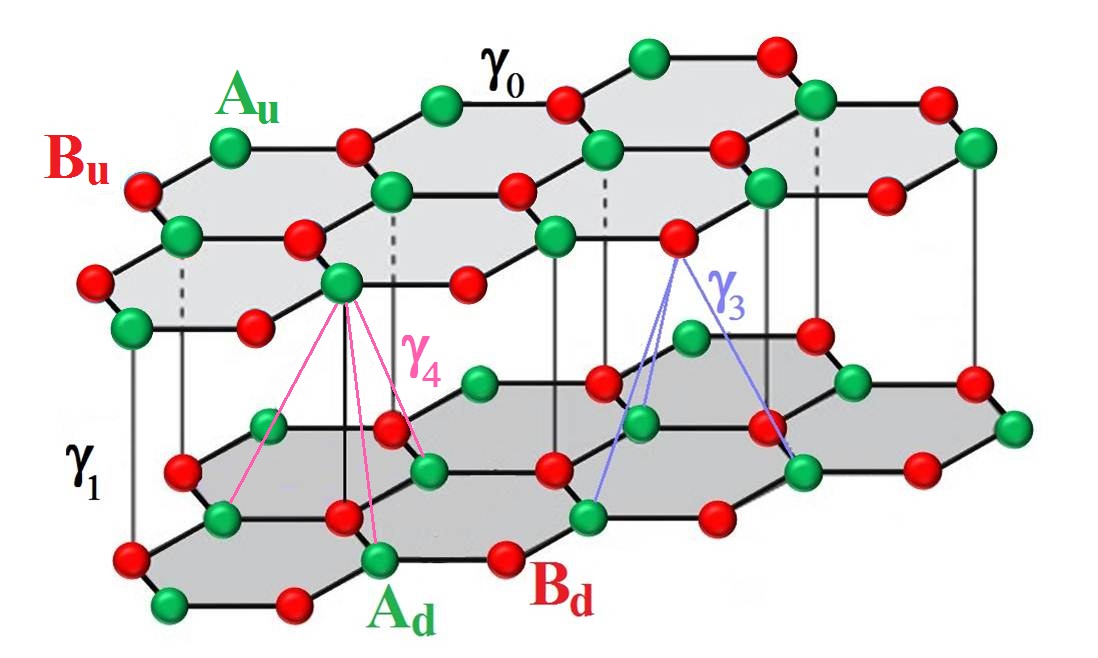}
\caption{Schematic view of bilayer graphene with Bernal stacking. Sub-lattices for the bottom layer ($ A_{d}, B_{d}$) and for the topper layer ($ A_{u}, B_{u}$) are coupled to each other with two skew interlayer couplings, $\gamma_3$ and $\gamma_4$.}
\label{fig1}
\end{figure}
Regarding to the importance of TrW, we derive an effective $2 \times 2$ Hamiltonian which describes low energy massive Dirac fermions in a bilayer graphene when Fermi velocity of charge carriers itinerant in each layer are different. By analysing term by term of the Hamiltonian, the influence of velocity asymmetry on the TrW deformation is investigated while an interlayer electric field is present. Emerging an indirect band gap when a small vertical bias is applied, can have significant effect on conductance through bilayer graphene. Moreover, it is proved that such interlayer asymmetry in Fermi velocity is able to generate the electron-hole (e-h) asymmetry in the spectrum.

This paper is organized as follows. After the introduction, in Sec.\ref{Sec1}, the effective Hamiltonian is derived by projecting the full Hamiltonian on the non-dimer sites. All results are presented in Sec.\ref{Sec2} which includes the four subsections. Trigonal warping modification induced by Fermi velocity asymmetry is investigated during Sec.\ref{Sub1} when no external bias is present. By applying an external bias, the spectrum and e-h asymmetry are studied in Sec.\ref{Sub2}. In Sec.\ref{Sub3}, the influence of the weaker skew interlayer coupling ($\gamma_4$) is shown on the e-h asymmetry. Finally, a general form of the spectrum which results in the indirect band gap is analysed in the presence of all terms of the effective Hamiltonian. At the end, Sec.\ref{Conclusion} is devoted to the conclusions.
\section{Effective $2 \times 2$ Hamiltonian}\label{Sec1}
A bilayer graphene with Bernal stacking is modelled such that two monolayer graphene are coupled to each other by using the dimer sites ($A_u-B_d$) coupling $\gamma_1$ and the skew interlayer coupling $\gamma_3$ between the non-dimer sites ($A_d-B_u$). The $A$ and $B$ sublattices in the upper layer have an additional interlayer skew hopping energy $\gamma_4$ with their correspondence sublattices in the lower layer ($A_{u}-A_{d}$ and $B_{u}-B_{d}$ couplings). The tight-binding Hamiltonian describing the band structure near the Dirac points was presented as the $4 \times 4$ matrix in the basis of the atomic orbitals sorted as $A_{d}, B_{u}, A_{u}, B_{d}$ in the unit cell\cite{mccann2013}. Here, since the Fermi velocity in each layer could be altered inducing by many factors such as strain, defect scatterings and so on, the following Hamiltonian is proposed based on its original one\cite{mccann2013} but with different Fermi velocities on each layer.

\begin{equation}
H=\begin{pmatrix}
-u/2&v_{3}\pi & -v_{4}\pi^{\dagger} &v_{d}\pi^{\dagger}\\
v_{3}\pi^{\dagger}&u/2&v_{u}\pi& -v_{4}\pi\\
-v_{4}\pi&v_{u}\pi^{\dagger}&u/2&\gamma_{1}\\
v_{d}\pi&-v_{4}\pi^{\dagger}&\gamma_{1}&-u/2
\end{pmatrix}
\label{Hamiltonian}
\end{equation}
where $v_u$ and $v_d$ are assumed to be the Fermi velocity of electrons in the upper and downer layers which might effectively differ from the velocity measured in pristine graphene. Here $u$ is the interlayer potential difference in the on-site energies:
$ \varepsilon_{A_u}=\varepsilon_{B_u}=u/2 $ and $ \varepsilon_{A_d}=\varepsilon_{B_d}=-u/2 $. In addition, two velocities are defined in based on the interlayer couplings as $v_{3}=\dfrac{\sqrt{3}}{2\hbar}a \gamma_{3}$ and $v_{4}=\dfrac{\sqrt{3}}{2\hbar}a \gamma_{4}$. Here $ \pi=-i\partial_{x}+\partial_{y} $. An effective $ 2\times 2 $ Hamiltonian is derived around the Dirac point at low energy limit which describes the spectrum of those orbitals belonging to the non-dimer sites $A_d-B_u$. The above Hamiltonian is divided into four blocks as the diagonal blocks $ h_{11}=-u/2\sigma_{z}+v_{3}(\sigma_{x}p_{x}-\sigma_{y}p_{y}) $,
$ h_{22}=u/2\sigma_{z}+\gamma_{1}\sigma_{x} $ and off diagonal blocks,
$$ h_{12}=\begin{pmatrix}
-v_{4}\pi^{\dagger} &v_{d}\pi^{\dagger}\\v_{u}\pi&-v_{4}\pi
\end{pmatrix}, \,\,\,\,\,\,h_{21}=\begin{pmatrix}
-v_{4}\pi&v_{u}\pi^{\dagger}\\v_{d}\pi&-v_{4}\pi^{\dagger}
\end{pmatrix}.
$$
\begin{figure}
\includegraphics[width=7cm]{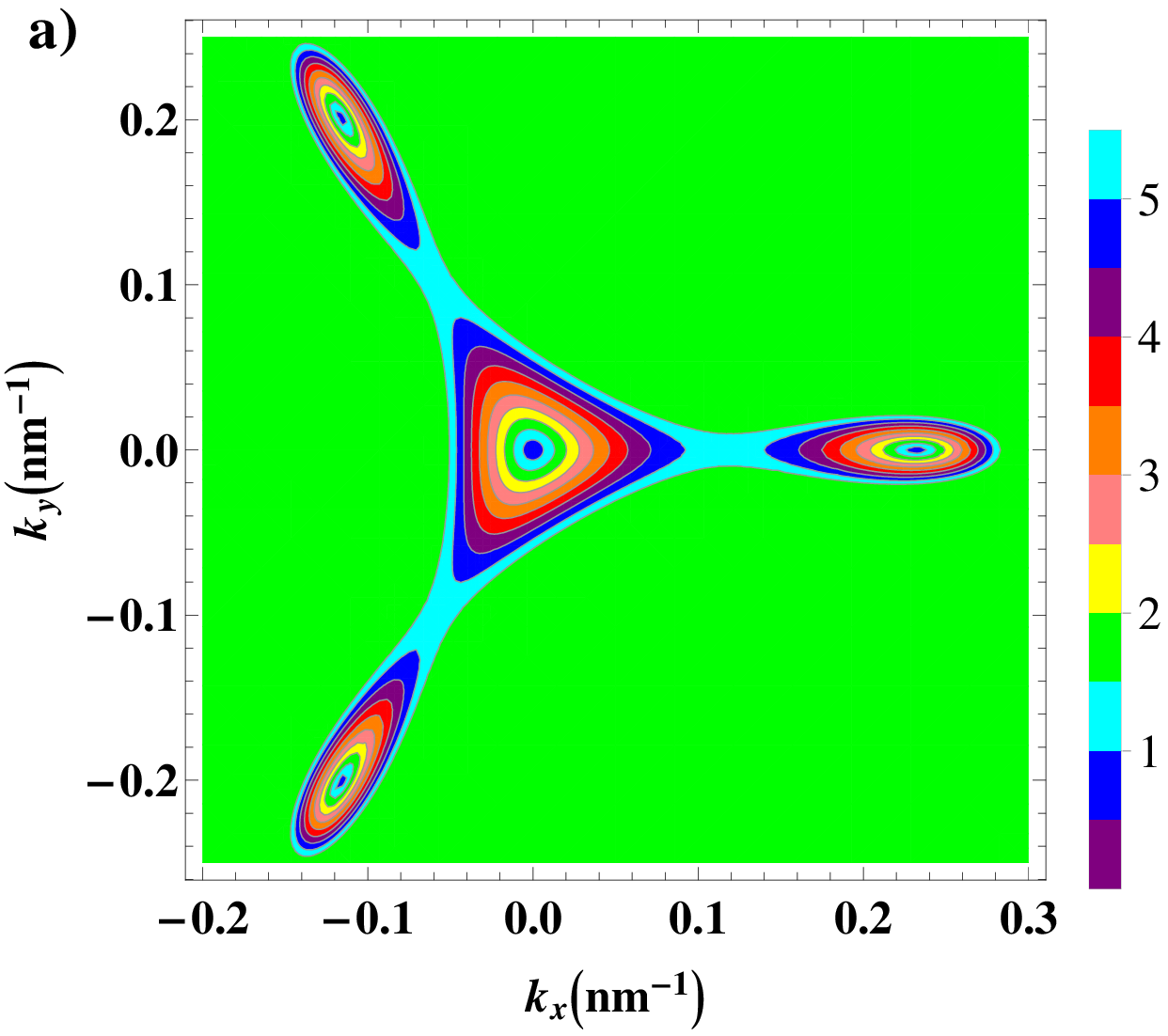}
 \includegraphics[width=7cm]{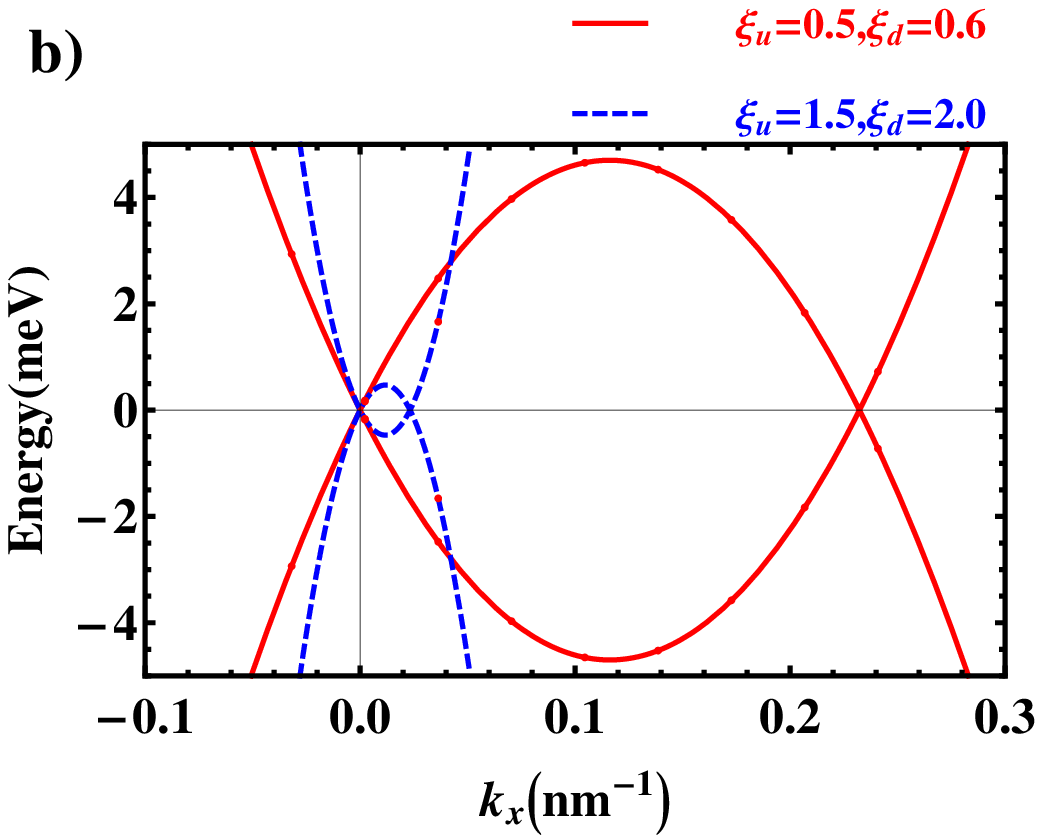}
\caption{ a) Contour plot of spectrum showing trigonal warping for different Fermi velocities of electrons in the up and down  layers, $ \xi_{u}=0.5, \xi_{d}=0.6$ and $\gamma_{3}=380 meV$. b) the band structure along the $k_x$ axis for two values of Fermi velocities, $ \xi_{u}, \xi_{d} $. Here, there is no gate bias. The weaker skew interlayer coupling is also considered to be zero, $\gamma_{4}=0$.}
\label{fig2}
\end{figure}
The eigenvalue equation of the above Hamiltonian is written as,
\begin{equation}
 \begin{pmatrix}
h_{11}&h_{12}\\h_{21}&h_{22}
\end{pmatrix}
\begin{pmatrix}
\alpha\\\beta
\end{pmatrix}
=\varepsilon \begin{pmatrix}
\alpha\\\beta
\end{pmatrix}
\end{equation}
where the dimer orbitals can be extracted in terms of the non-dimer orbitals as the following $ \beta=(\varepsilon-h_{22})^{-1}h_{21}\alpha $. In fact, to obtain projected Hamiltonian on the non-dimer sites, those terms which are related to the dimer sites must be finally eliminated at low energies. This method is presented in Ref.[\onlinecite{mccann2013}]. By substitution of $\beta$, the eigenvalue equation is approximated in low energy limit as
\begin{equation}
[h_{11}-h_{12}h_{22}^{-1}h_{21}]\alpha\approx \varepsilon S \alpha
\end{equation}
where $ S=1+h_{12}h_{22}^{-2}h_{21}$. By defining a transformation like as $$ \alpha=S^{-1/2} \varphi $$, the effective Hamiltonian is read as
\begin{equation}
H^{eff}\approx S^{-1/2}[h_{11}-h_{12}h_{22}^{-1}h_{21}]S^{-1/2},
\end{equation}
 where
\begin{equation}
S=1+X,\,\,\,\,\,\,X=h_{12}h_{22}^{-2}h_{21}.
\end{equation}
\begin{figure}
\centering
\includegraphics[width=7cm]{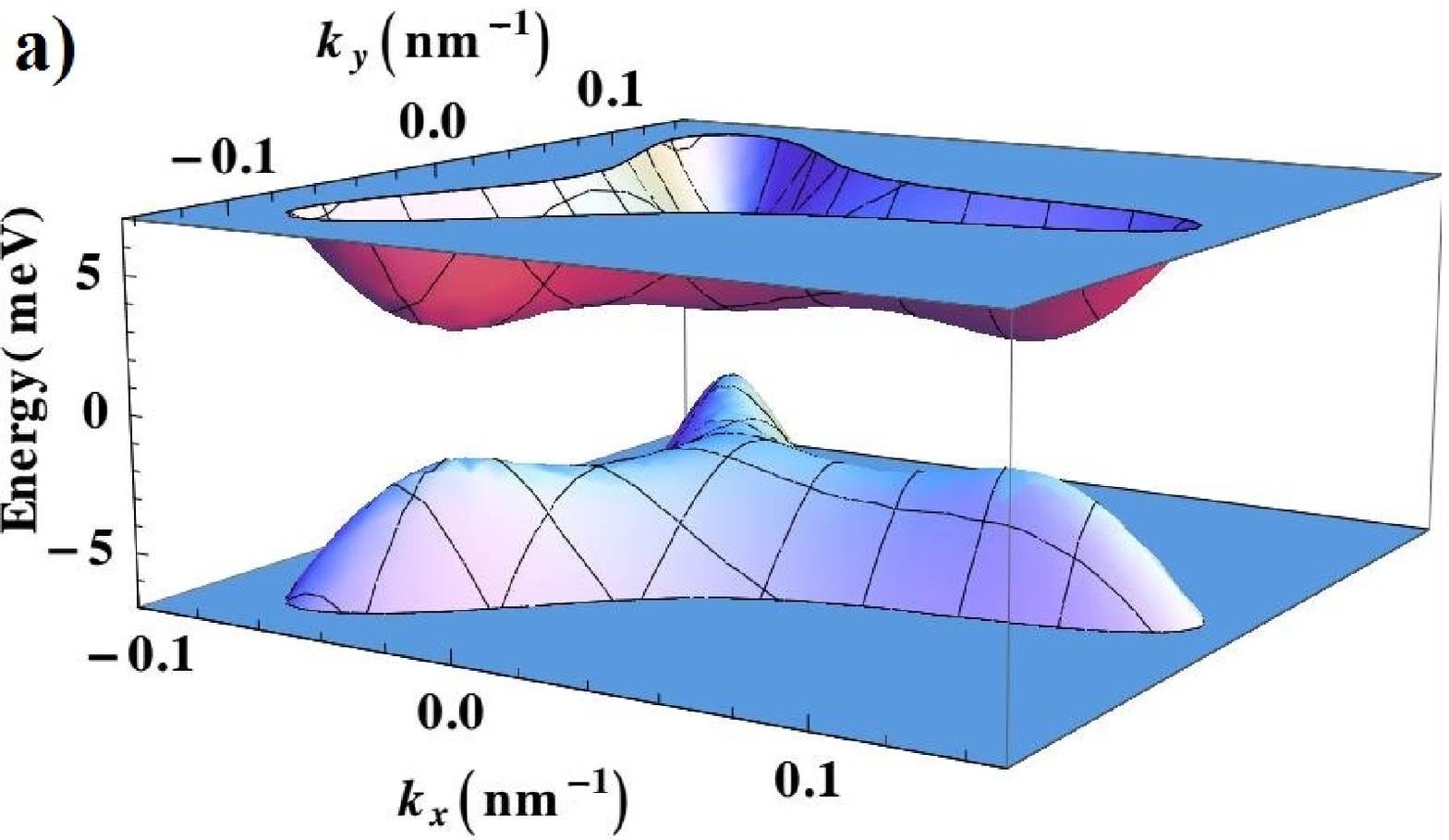}
\includegraphics[width=7cm]{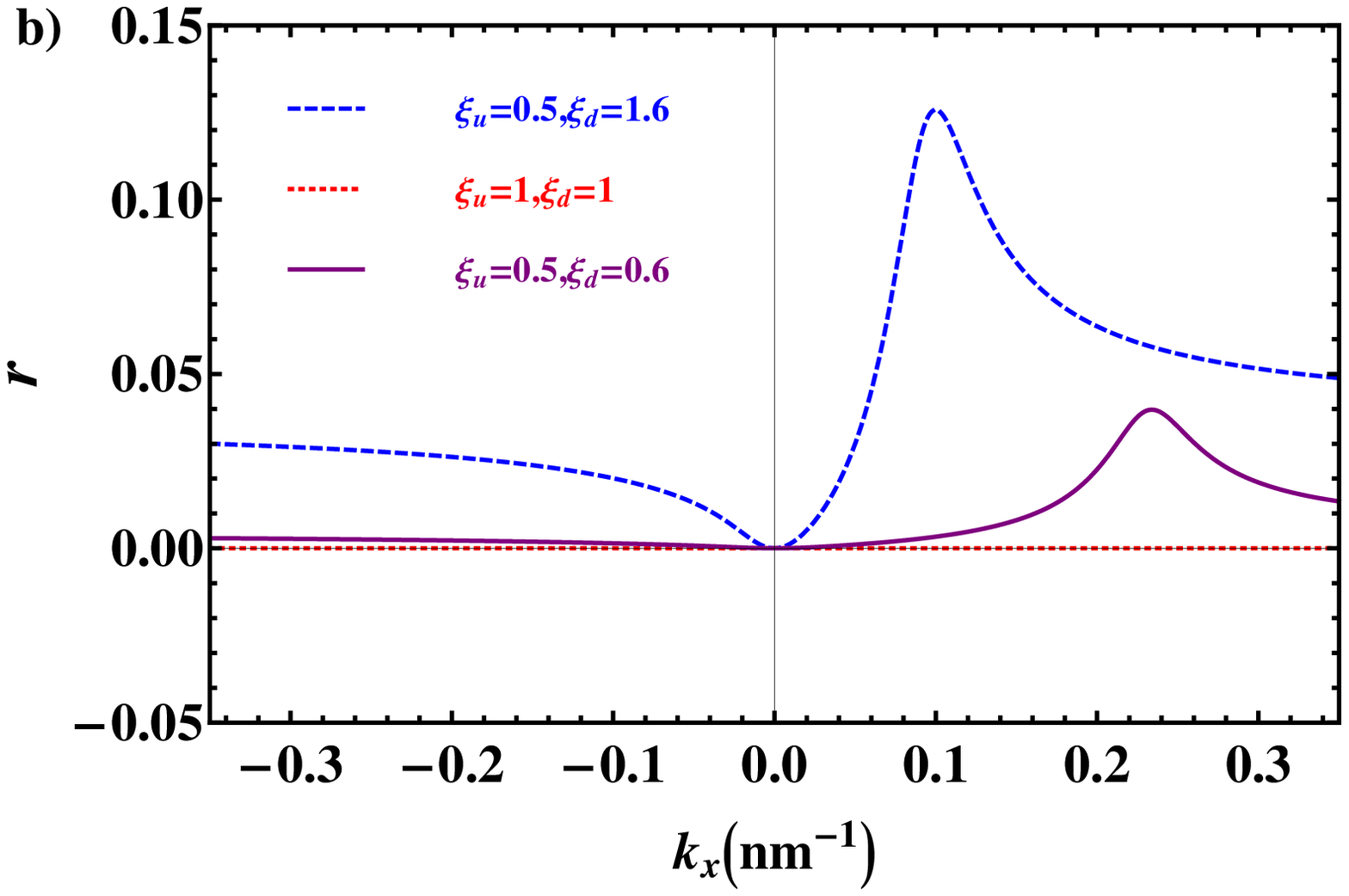}
\caption{ a)the band structure, b) the e-h asymmetric factor $r$ in terms of $ k_{x} $ under application of small perpendicular bias $ u=5 meV$.}
\label{fig3}
\end{figure}
For small values of $X$, we have $ S^{-1/2} \approx 1-X/2 $. In these approximations, the system parameters such as  $ |\gamma_{3}|, |\gamma_{4}|, |u|$ are too small in compared to $\gamma_0$. Therefore, in our calculations, these parameters are kept upto the linear term while the momentum is kept upto the quadratic term. As a result, the effective Hamiltonian is derived as the following,
\begin{equation}
\hat{H}_{eff}=\hat{h}_{0}+\hat{h}_{w}+\hat{h}_{u}+\hat{h}_{4}
\end{equation}
where
\begin{equation}
\begin{aligned}
&\hat{h}_{0}=-\dfrac{v_{u}v_{d}}{\gamma_{1}}\begin{pmatrix}
0& \pi^{\dagger^{2}}\\
\pi^{2}&0
\end{pmatrix}\\
&\hat{h}_{w}=v_{3}\begin{pmatrix}
0&\pi\\ \pi^{\dagger}&0
\end{pmatrix}\\
&\hat{h}_{u}=-u/2 \begin{pmatrix}
1&0\\0&-1
\end{pmatrix}
+\dfrac{u}{\gamma_{1}^{2}}\begin{pmatrix}
v_{d}^{2}\pi^{\dagger}\pi&0 \\
0&-v_{u}^{2}\pi\pi^{\dagger}
\end{pmatrix}\\
&\hat{h}_{4}=\dfrac{2v_{4}}{\gamma_{1}}\begin{pmatrix}
v_{d}\pi^{\dagger}\pi&0 \\
0&v_{u}\pi\pi^{\dagger}
\end{pmatrix}.
\end{aligned} \label{effective_hamil}
\end{equation}
To check our results, we derived exactly the same form for the effective Hamiltonian by using another equivalent method, namely as the greens function method. This method is presented in the Appendix \ref{Green_method}. The Lowdin partitioning method\cite{Lowdin} is also a general and powerful method which could be applied to derive this effective Hamiltonian. This method also concludes the same results for the effective Hamiltonian as proved in Appendix \ref{Lowdin_method}.

As a conclusion of the first term $\hat{h}_0$, the effective mass of chiral electrons in the parabolic spectrum depends on Fermi velocities in the up and down layers, $m=\gamma_1/ 2 v_u v_d$. In this Hamiltonian, $\hat{h}_{w}$ is responsible for TrWs while $\hat{h}_{u}$ is denoted to the applied bias. $\hat{h}_{4}$ shows the influence of the skew interlayer coupling $\gamma_4$ on the spectrum. Therefore,the general relation for the dispersion of the effective Hamiltonian is derived as the following formula,
\begin{equation}
\begin{aligned}
&E_{\pm}=\frac{1}{2\gamma_{1}^{2}} \lbrace k^{2}[uv_{F}^{2}(\xi_{d}^{2}-\xi_{u}^{2})+2 v_{4}v_{F}(\xi_{u}+\xi_{d})\gamma_{1}] \\
&\pm \sqrt{k^{4}x^{2}+4k^{2}v_{3}^{2}\gamma_{1}^{4}-4k^{3}v_{3}\gamma_{1}^{2}x \cos (3\varphi)+(u\gamma_{1}^{2}-k^{2}f)^{2}} \ \rbrace,\\
&x=uv_{4}v_{F}(\xi_{u}-\xi_{d})+2v_{F}^{2}\xi_{d}\xi_{u}\gamma_{1},\\
&f=uv_{F}^{2}(\xi_{u}^{2}+\xi_{d}^{2})+2v_{4}v_{F}(\xi_{d}-\xi_{u})\gamma_{1}.
\end{aligned}
\label{E-k}
\end{equation}
where $\xi_{u}=v_u/v_{f}, \xi_{d}=v_d/v_{f}$. It should be noted that $\hat{h}_{w}$ has also the same momentum dependence as $\hat{h}_0$. However, it just gives a small additional contribution to the effective mass $m$. So we skip the quadratic term of momentum in $\hat{h}_{w}$.

\section{Results}\label{Sec2}
In what follows, we investigate the effect of velocity asymmetry on the dispersion of the effective Hamiltonian in some special cases.
\subsection{Trigonal Warping dependence on velocity asymmetry with no external bias} \label{Sub1}
Let us first focus on the gapless case in which external gate bias is absent $u=0$ and the weaker skew coupling $\gamma_4$ is zero. So only two terms are remained in the effective Hamiltonian shown by Eq.\ref{effective_hamil} as $$H_{eff}=\hat{h}_{w}+\hat{h}_{0}$$. There is a triangular perturbation to the circular iso-energetic lines known as TrW which is described by $\hat{h}_w$ in the effective Hamiltonian. It is interesting that at a fast look, there is no trace of the velocity asymmetry in the term which is responsible for TrW, $\hat{h}_w$. However, as we will prove later, TrW can be changed by inducing an asymmetry in Fermi velocity. The energy dispersion arising from the effective Hamiltonian is given by
\begin{equation}
E_{\pm}=\pm\dfrac{1}{\gamma_{1}}\sqrt{v_{f}^{4}\xi_{u}^{2}\xi_{d}^{2}k^{4}+v_{3}^{2}\gamma_{1}^{2}k^{2}
-2v_{f}^{2}v_{3}\xi_{u}\xi_{d}\gamma_{1} k^{3}\cos(3\varphi)}.
\label{Ew}
\end{equation}
 The specific momentum of the leg pocket can be obtained as
\begin{equation}
k_{w}=\dfrac{v_{3}\gamma_{1}}{4v_{f}^{2}\xi_{d}\xi_{u}}[3\cos(3\varphi)+\sqrt{9\cos^{2}(3\varphi)-8}]\label{Kw}.
\end{equation}
\begin{figure}
\centering
\includegraphics[width=7cm]{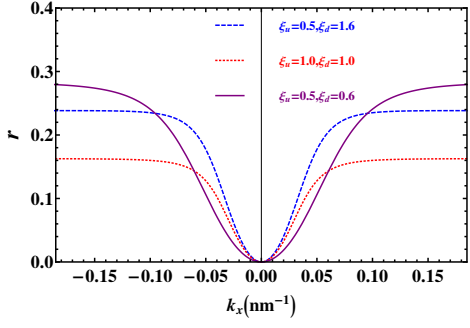}
\caption{The electron-hole asymmetric factor $r$ in terms of $k_x$ for the case of $\gamma_{3}=0$ and $ u=5 meV$.}
\label{fig4}
\end{figure}

For $ \varphi=0,2\pi/3,4\pi/3 $, the momentum $k_{w}$ attributed to the three leg pockets is equal for all three legs meaning that velocity asymmetry does not change three fold rotational symmetry of TrW at low energies. This symmetry is confirmed in Fig.\ref{fig2}a which shows a contour plot of the conduction band as a function of $k_{x}, k_{y}$. As a result of Eq. \ref{Kw}, $k_w$ is inversely proportional to the multiplication of the up and down velocities, $\xi_u \times \xi_d$. As shown in Figure \ref{fig2}b, with a decrease in velocities multiplication, correspondingly, $ k_{w} $ is also increasing and emerging away from the Dirac point. Moreover, it should be noted that the velocity asymmetry preserves the e-h symmetry provided that the weaker skew interlayer coupling $\gamma_4$ is absent.
\begin{figure}
\centering
\includegraphics[width=7cm]{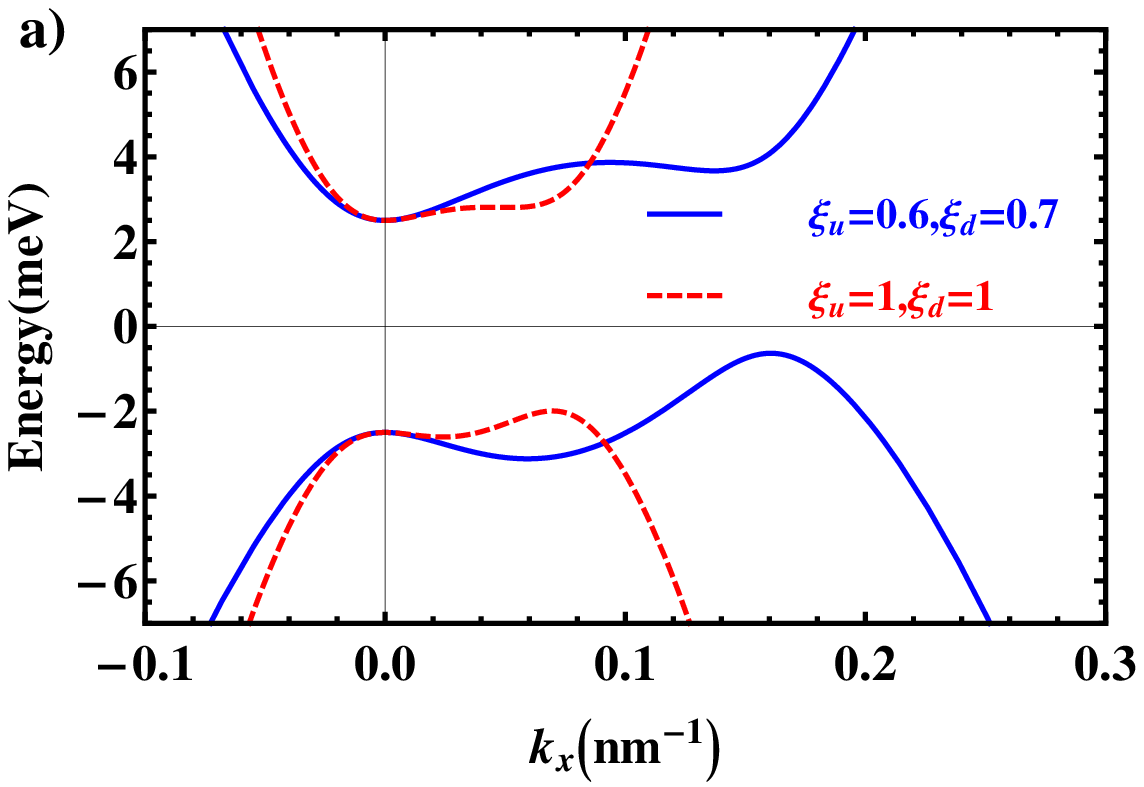}
\includegraphics[width=7cm]{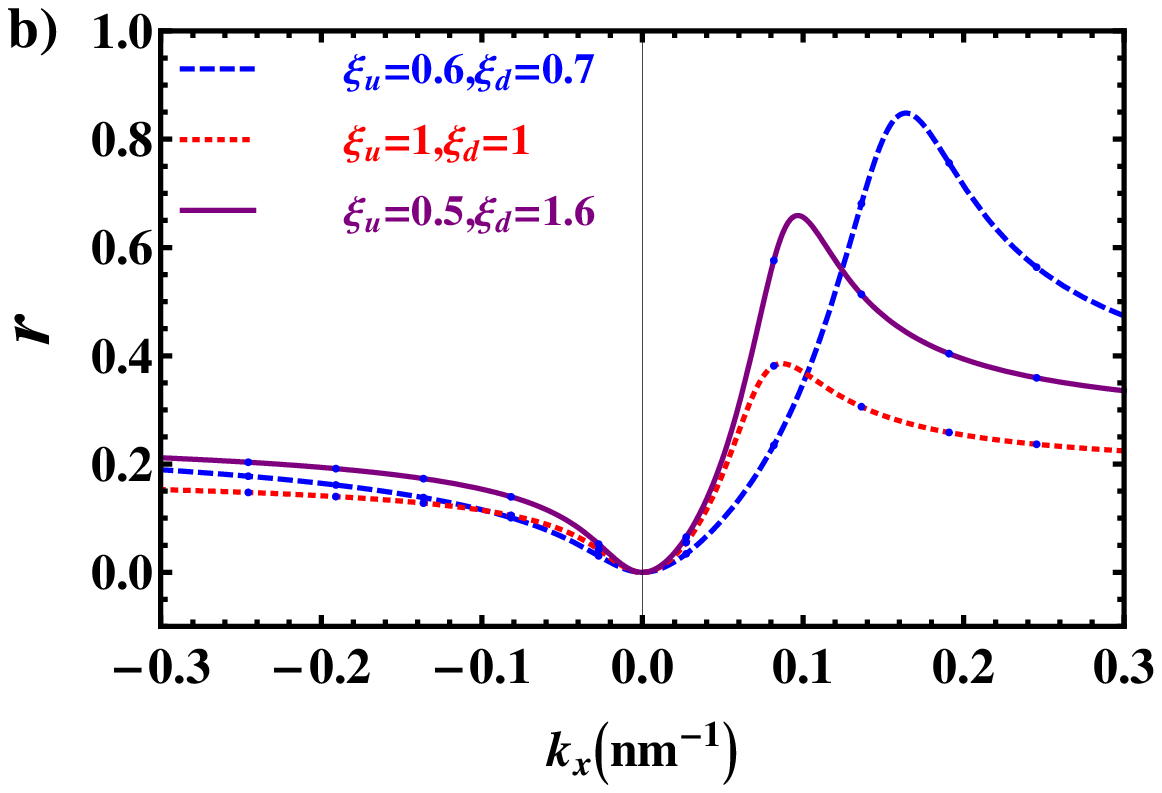}
\caption{a) The band structure of bilayer graphene in low energy limit when Fermi velocity on each layer is different from the velocity of electrons in pristine graphene as $v_{u}=0.5 v_F, v_{d}=0.6 v_F $, b) The e-h asymmetric factor $r$ as a function of $k_{x}$. In this spectrum, skew couplings are considered to be the same as values reported in reference \onlinecite{Kuzmenko2009} of Table.\ref{table}. Small external bias is also applied, $u=5 meV$.}
\label{fig5}
\end{figure}
\subsection{Spectrum dependence on velocity asymmetry in presence of external bias} \label{Sub2}
In the next step, let us look at the gapped spectrum when an interlayer potential asymmetry $ (\hat{h}_{u}) $ is applied perpendicular to bilayer graphene. So at low external biases in compared to $\gamma_1$, $ \hat{H}_{eff} $ is written as:
\begin{equation}
\hat{H}_{eff}=\hat{h}_{0}+\hat{h}_{w}+\hat{h}_{u}
\end{equation}
Still it is supposed $\gamma_4=0$. In this case, the energy spectrum is extracted from the effective Hamiltonian of Eq. \ref{E-k} as the following.
\begin{equation}
\begin{aligned}
& E_{\pm}=\dfrac{k^{2}uv_{f}^{2}(\xi_{d}^{2}-\xi_{u}^{2})}{2 \gamma_{1}^{2}} \pm \frac{1}{2}\\
&[\dfrac{k^{4}u^{2}v_{f}^{4}(\xi_{d}^{2}+\xi_{u}^{2})^{2}}{\gamma_{1}^{4}}-
\dfrac{2k^{2}u^{2}v_{f}^{2}(\xi_{d}^{2}+\xi_{u}^{2})}{\gamma_{1}^{2}}+\dfrac{4k^{4}v_{f}^{4}\xi_{d}^{2}\xi_{u}^{2}}{\gamma_{1}^{2}}\\
&-\dfrac{8k^{3}v_{f}^{2}\xi_{d}\xi_{u} v_{3}\cos(3\varphi)}{\gamma_{1}}+u^{2}
+4k^{2}v_{3}^{2}]^{1/2}.
\end{aligned}
\label{Egap}
\end{equation}
\begin{figure}
\centering
\includegraphics[width=6cm]{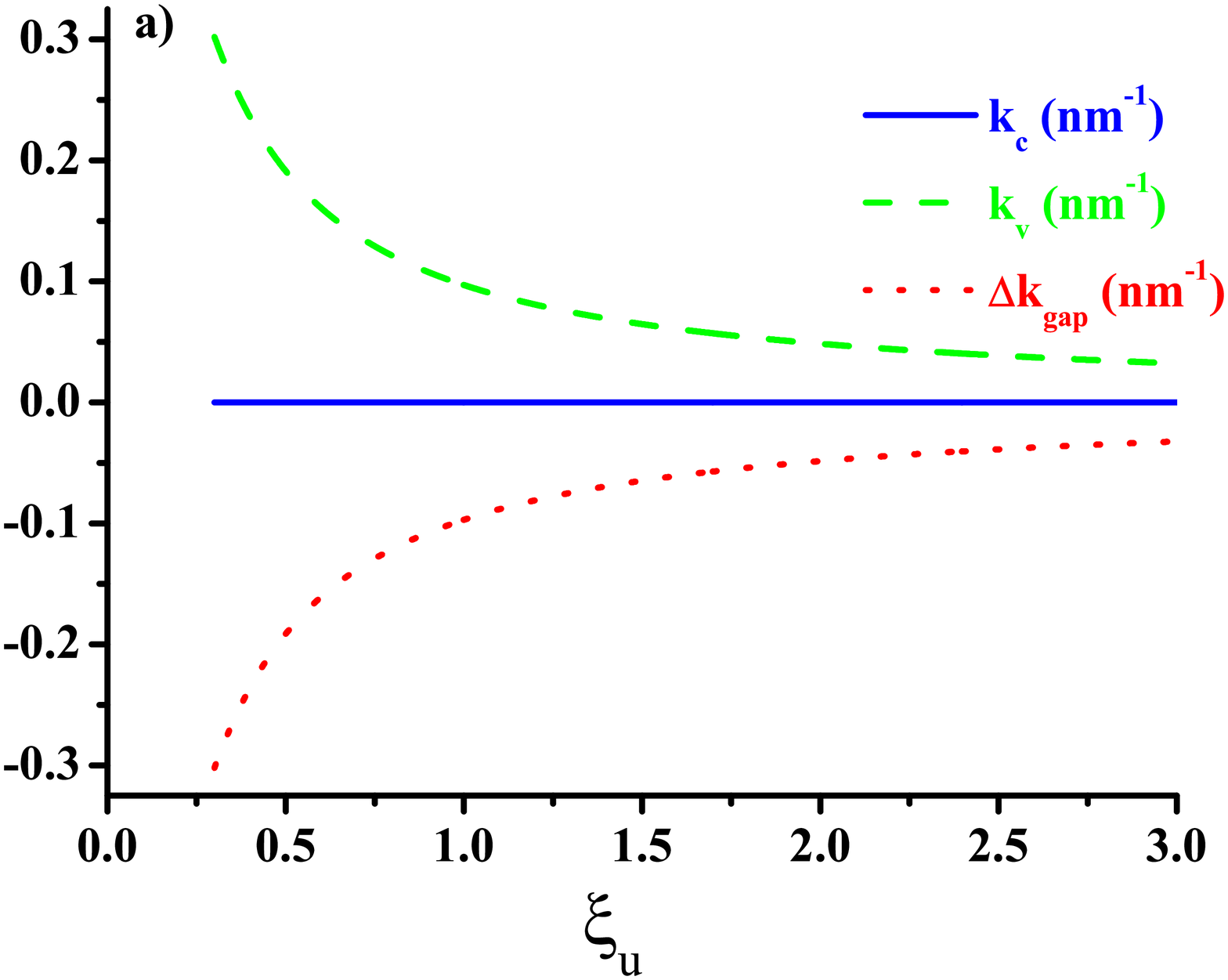}
\includegraphics[width=6cm]{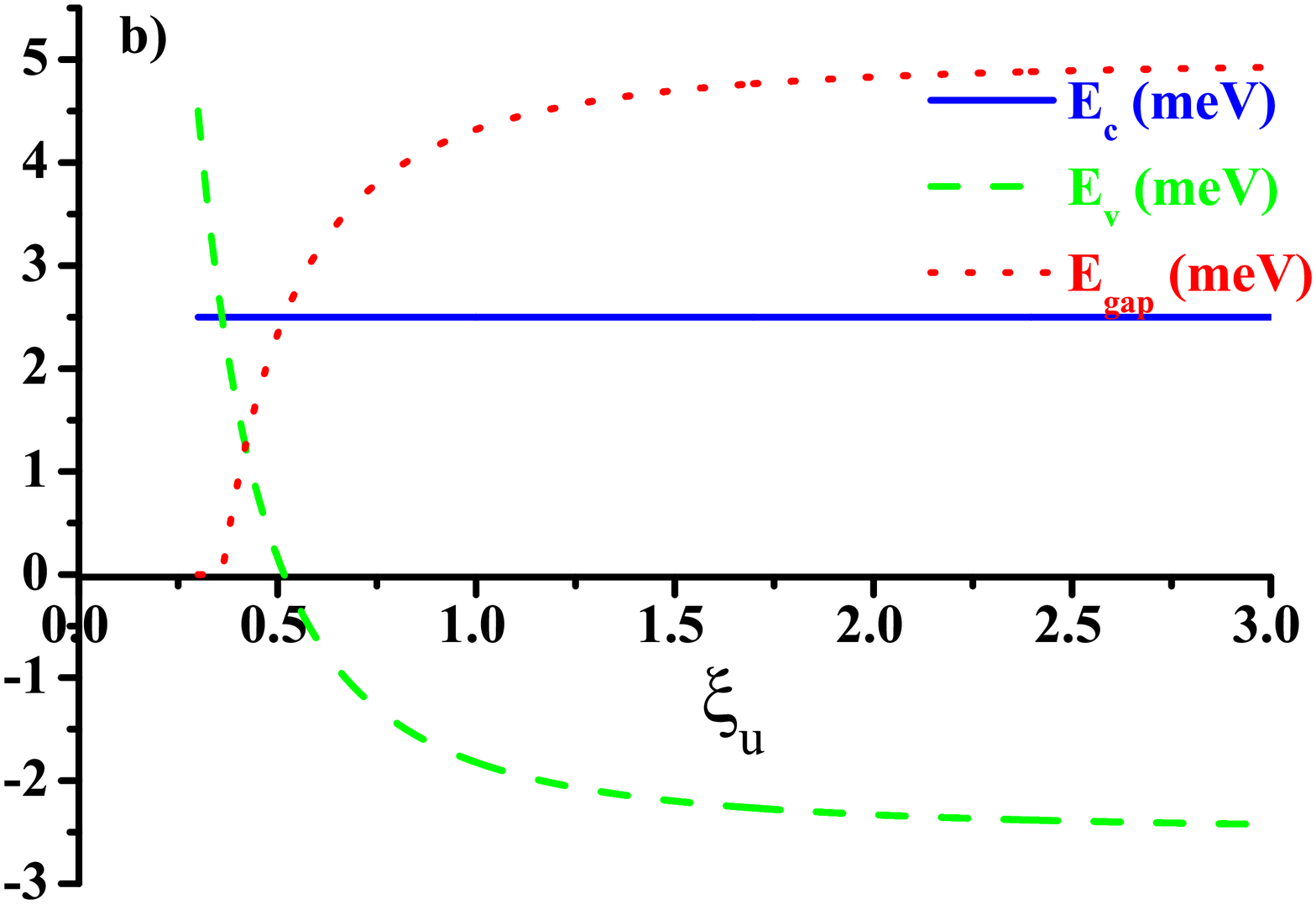}
\caption{ a) The momentum attributed to the conduction and valence band edges $k_c$ and $k_v$ as a function of Fermi velocity in the upper layer $\xi_u$ when Fermi velocity in the bottom layer is fixed to $\xi_d=0.7$. $\Delta k_{gap}=k_c-k_v$ indicates the amount of indirectness of the band gap. b) The energy gap and also the conduction and valence band edges ($E_c,E_v$) dependence on Fermi velocity of the upper layer. The skew couplings are considered to be the same as values reported in refrence \onlinecite{Kuzmenko2009} of Table.\ref{table}. The external potential is as small as $ u=5 meV$.}
\label{fig6}
\end{figure}
The above equation clearly shows that the velocity asymmetry accompanied by the applied bias breaks the e-h symmetry. However, when $\xi_u=\xi_d$, the e-h symmetry is preserved. Figure \ref{fig3}a represents the band structure of the gapped spectrum for $ \xi_{u}=0.5, \xi_{d}=1.6$ in presence of small external bias, $ u=5 meV $. Let us define the parameter $r$ to measure breaking of the e-h asymmetry.
\begin{equation}
 r=(|E_{+}|-|E_{-}|)/|E_{+}|.
\end{equation}
The e-h asymmetric factor $r$ varies between $0$, for the full e-h symmetric case and $1$, for the full asymmetric case. The e-h asymmetric factor depends on the momentum. Fig.\ref{fig3}b shows this asymmetric factor in terms of $k_x$ for different values of velocity in each layer. It is clear that e-h asymmetry is increased around the pockets of three legs. The parameter $ r $ is proportional to $\dfrac{\xi_{d}^{2}-\xi_{u}^{2}}{\xi_{d}^{2}\xi_{u}^{2}}$ which is also demonstrated in Fig. \ref{fig3}b. It should be insisted that this asymmetry is present only if an asymmetric velocity is present. Moreover, in this case, the band gap is direct and independent of the velocity asymmetry.

\subsection{Skew interlayer coupling effect, $\gamma_4$: The electron-hole asymmetry} \label{Sub3}
To clarify the effect of $\gamma_4$, the spectrum of Eq. \ref{E-k} is studied by supposing $\gamma_{3}=0$. So in this case, the
effective Hamiltonian and its spectrum are written as
\begin{equation}
\hat{H}_{eff}=\hat{h}_{0}+\hat{h}_{4}+\hat{h}_{u}.
\end{equation}

The presence of $\gamma_{4}$ can break the e-h symmetry even if $\xi_u=\xi_d$. However, an other factor which can break the e-h symmetry is the asymmetry in velocities. Figure \ref{fig4} shows the e-h asymmetric factor $r$ as a function of $ k_{x}$ for different values of the up and down velocities. $r$ behaves as $k^2$ close to Dirac point as the following
\begin{equation}
\begin{aligned}
&r \simeq \frac{1}{2u\gamma_{1}^{2}}[uv_{F}^{2}(\xi_{d}^{2}-\xi_{u}^{2})+2 v_{4}v_{F}(\xi_{u}+\xi_{d})\gamma_{1}] k^{2}.
\end{aligned}
\end{equation}
It is seen that $r$ is non-zero for the case of $\xi_u=\xi_d=1$. In the absence of TrW, the band structure of massive Dirac fermions which keeps its 'Mexician hat' shape, has a circular symmetry around the Dirac points.
\subsection{ Velocity asymmetry and Spectrum: Indirect band gap} \label{Sub4}
Finally, let us assume both interlayer couplings $ \gamma_{3}$ and $\gamma_{4} $ to be present in the effective Hamiltonian. So in this general case, we consider all terms, $$\hat{H}_{eff}=\hat{h}_{0}+\hat{h}_{w}+\hat{h}_{u}+\hat{h}_{4}.$$
The spectrum of this general Hamiltonian was presented in Eq.\ref{E-k}. In this section, it is demonstrated that considering all terms in the effective Hamiltonian can lead to a significant impact on the band structure. The band structure in Fig. \ref{fig5}a which is plotted for the two different cases shows that the band gap is indirect for both symmetric and asymmetric velocities on each layers. Moreover, large deviation from the e-h symmetry is obvious in this case. Fig. \ref{fig5}b confirms that for both $\xi_u$ and $\xi_d$ to be lesser than unity, the factor $r$ reaches to nearly unity at the leg pockets meaning that a full e-h asymmetry emerges in these points.

In what follows, energy gap and it's indirect behaviour is investigated when asymmetry in Fermi velocities is applied. To evaluate the band gap, the conduction and valence band edges $ E_{c}$ and $E_{v}$ and their attributed momentums $k_{c}$ and $k_{v}$ should be numerically studied as a function of $\xi$. This is performed in Figs. \ref{fig6}a,b. As a result, the conduction band edge is pinned to $E_{c}=0$ while the valence band edge $E_v$ emerges in the leg pockets. This behaviour was also shown in Fig. \ref{fig5}a. The momentum attributed to the valence band edge $k_v$ which occurs at the trigonal pockets, decreases with $\xi_u$ as indicated in Fig. \ref{fig6}a. In this figure, $ \Delta k_{gap}=k_{c}-k_{v}$ represents the amount of indirectness of the band gap which is increased for $\xi<1$. What is important here is that even if Fermi velocities are equal in the two layers, $\xi_u=\xi_d$, such indirect band gap is still present, while if we do not consider the skew hopping integrals, the band gap is direct in the symmetric Fermi velocities on layers\cite{cheraghchi2014}. In fact, the skew interlayer couplings can result in an indirect band gap even in the symmetric case of Fermi velocity. Furthermore, the energy gap $E_{gap}=E_c-E_v$ which is plotted in Fig. \ref{fig6}b, increases as a function of $\xi_{u}$ and finally saturates to the perpendicular applied bias, $u$, for the velocities greater than Fermi velocity of pristine graphene. The energy gap tends to be zero in very small Fermi velocities.
\section{Experimental Relevance}
Recently, Raman spectroscopy of bilayer graphene revealed that local strain fluctuations which emerge during the growth or annealing process, is stronger at the bottom layer which in contact with the substrate in compared to the top layer which is only in connection with the bottom layer\cite{Verhagen2015}. This fact accompanied to the nanoscale ripples and curvatures\cite{Falko2013,Meng2013} and also scatterings arising from the defects and dopings\cite{Attaccalite2009} lead to an effective change in Fermi velocity on each layer. Additional to the strain and mismatching with the substrate\cite{Ortix2012,Yin2015}, remote electrostatic interaction induced by substrate as a dielectric screening\cite{Siegel2011,Jang2008,Ni2008} and also electron-electron interaction\cite{Trevisanutto2008,Elias2011,Bostwick2007} generate other types of scatterings which could result in Fermi velocity modification.

\begin{figure}
\centering
\includegraphics[width=7cm]{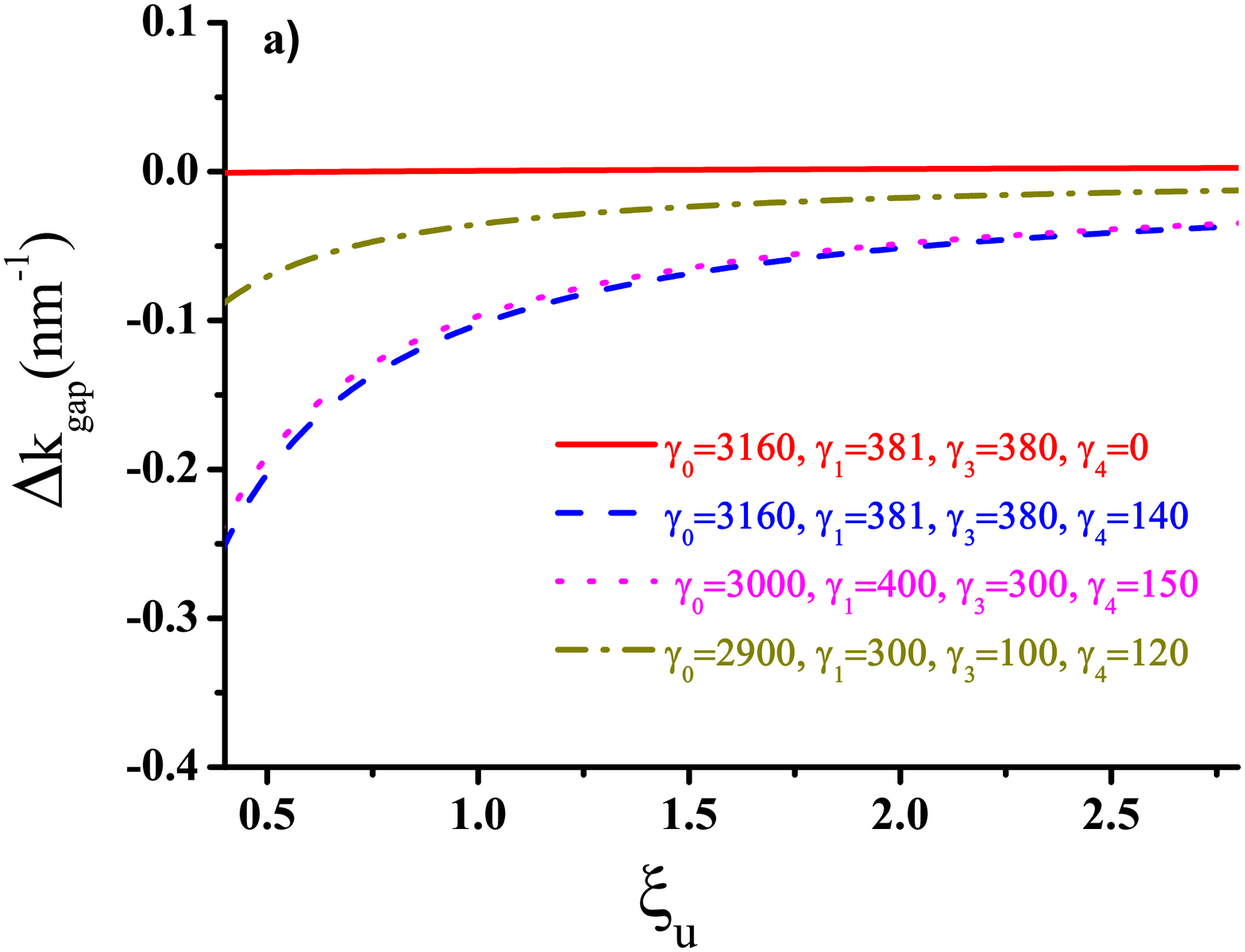}
\includegraphics[width=7cm]{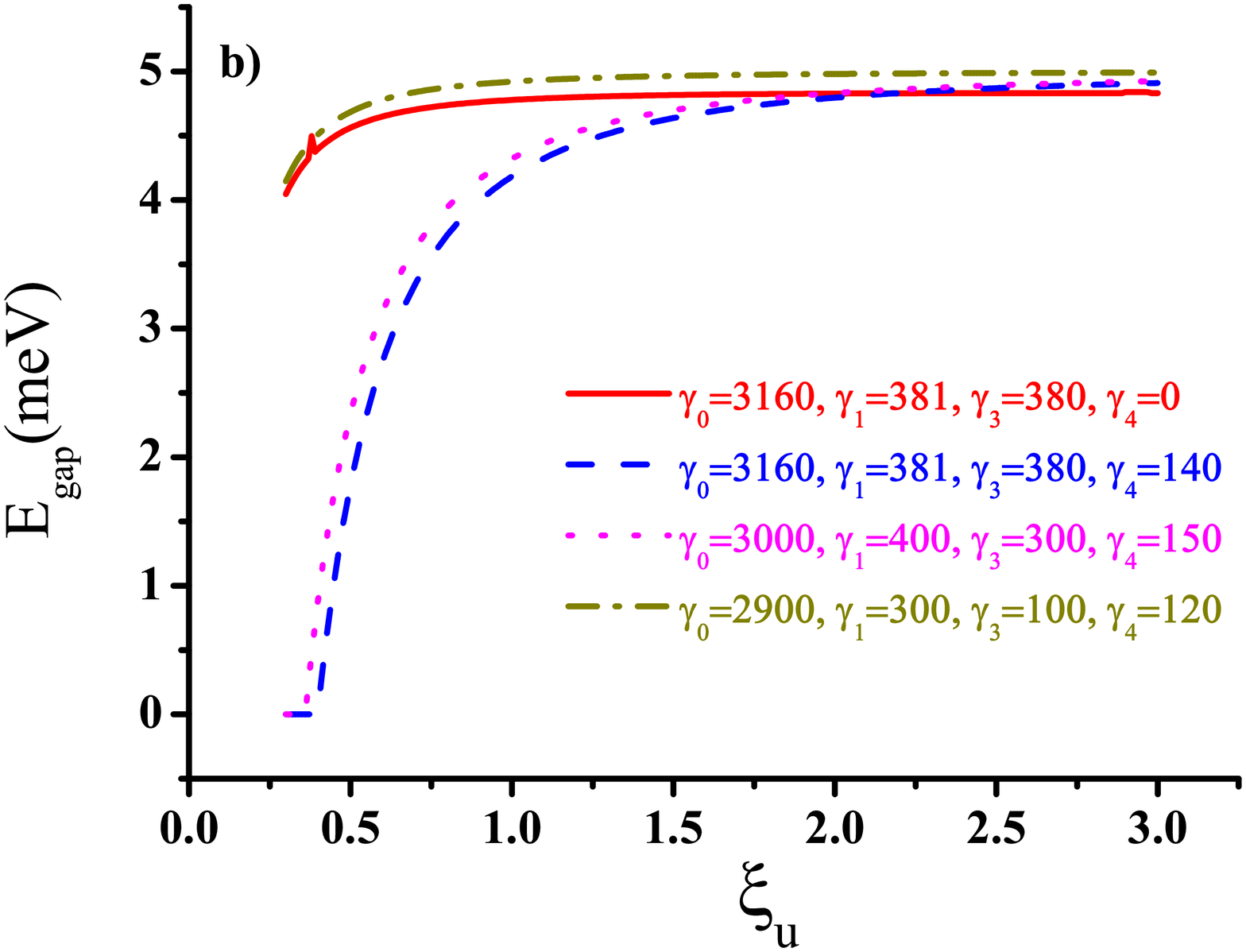}
\caption{ a) The momentum difference attributed to the conduction and valence band edges $\Delta k_{gap}=k_c-k_v$ and b) the energy gap in terms of Fermi velocity in the topper layer $\xi_u$ when Fermi velocity in the bottom layer is fixed to $\xi_d=0.7$. The skew couplings are considered to be the same as values reported in Table.\ref{table}. The external potential is as small as $ u=5 meV$.}
\label{fig7}
\end{figure}

On the other hand, a dramatically break of the electron-hole asymmetry has been recently observed experimentally in graphene monolayers which is originated to strain and charge-defect scatterings \cite{Bai2015}leading to different Fermi velocities for the electrons and holes. Besides, a significant electron-hole asymmetry was also experimentally observed in bilayer graphene by cyclotron resonance\cite{Henriksen2008} and infra-red spectroscopy\cite{Li2009}. These experimental evidences of a giant electron-hole asymmetry propose that one can investigate the influence of an asymmetry in Fermi velocity on the band structure especially on trigonal warping and Lifshitz transition. Since different measurements have been reported in literature, we have also checked robustness of the above results against different values of $ \gamma_{0},\gamma_{1}, \gamma_3$ and $\gamma_4$. Table \ref{table} lists some of the system parameters measuring by different groups. The results extracted from these measurements, are presented in Figures \ref{fig7}a,b which show $ \Delta k_{gap} $ and $ E_{gap} $ in terms of $\xi_u$. The red solid line in Fig. \ref{fig7}a demonstrates that the band gap is direct for the case of Sec. \ref{Sub1} in which $\gamma_4=0$. In Fig. \ref{fig7}, indirectness of the gap is investigated for those three measurements shown in Table. \ref{table}. As it is seen for all three cases, the band gap remains indirect and is tunable with the Fermi velocity engineering in each layer.
\begin{table}
\centering
 \begin{tabular}{|c c c c |}
 \hline
 Reference    & [\onlinecite{Kuzmenko2009}]   &[\onlinecite{Zhang2008}]  & [\onlinecite{Malard2007}] \\ [0.5ex]
 \hline\hline
 $ \gamma_{0} $ & 3160 & 3000 & 2900 \\
 \hline
 $\gamma_{1}$   & 381  & 400  & 300 \\
 \hline
 $\gamma_{3}$   & 380  & 300  & 100 \\
 \hline
 $\gamma_{4} $  & 140 & 150  & 120 \\
 \hline
\end{tabular}
\caption{The hopping integral energy (in meV) of intera and inter-layer couplings in different measurements.}
\label{table}
\end{table}
 \section{Conclusion}\label{Conclusion}
The band structure of bilayer graphene is modified when Fermi velocities of the up and down layers are different. Asymmetric Fermi velocity is originated from many factors such as curvatures or strain among the graphene sheet, electron-electron interaction and superlattice structures created by potentials or titled graphene sheets. In this paper, at low energy spectrum, modification of trigonal warping and also indirectness of the band gap are investigated. To perform this purpose, we develop an effective Hamiltonian at low energies in presence of velocity and potential asymmetry manufactured on the topper and lower layers. As a result of the spectrum, it is proved that the electron-hole symmetry breaks whenever Fermi velocities of electrons on layers are asymmetric.
\acknowledgments
H.C. thanks the International Center for Theoretical Physics (ICTP) for their hospitality and support during a visit in which part of this work was done.

\appendix
\section{Greens function method}\label{Green_method}
To derive the effective Hamiltonian, an alternative method is the Greens function method\cite{falko2006} to project the Hamiltonian on the non-dimer sites. The Greens function $G=(\varepsilon-H)^{-1}$ is defined as
\begin{equation}
G=\begin{pmatrix}
G_{11}&G_{12}\\G_{21}&G_{22}
\end{pmatrix}=\begin{pmatrix}
(G_{11}^{0})^{-1}&-H_{12}\\-H_{21}&(G_{22}^{0})^{-1}
\end{pmatrix}^{-1}.
\end{equation}
The projected Hamiltonian on the non-dimer sites is performed as the following $\tilde{H}_{11}= \varepsilon-G_{11}^{-1}=H_{11}+H_{12}G_{22}^{0}H_{21}$ where $ G_{22}^{0}=(\varepsilon-H_{22})^{-1}$.
In the low energy limit $ \vert \varepsilon \vert \ll \gamma_{1}$, it is possible to expand $G_{22}^{0} $ in terms of $ \gamma_{1}^{-1}$.
$$ G_{22}^{0}=(\varepsilon-H_{22})^{-1}=(\varepsilon-\frac{u}{2} \sigma_{z}-\gamma_{1}\sigma_{x})^{-1} $$
It is supposed that the following parameters $ |\gamma_{3}|, |\gamma_{4}|, |u|$ are too small in compared to $\gamma_0$ so that one can neglect their quadratic terms during the expansion. In addition, the expansion terms are kept upto the quadratic terms in the momentum.
\section{Lowdin Partitioning method}\label{Lowdin_method}
The Lowdin partitioning method which is based on the Schrieffer-Wolff transformation, is a general and powerful perturbative approach. This less known method is used when one can divide unperturbed states into two classes. Then the influence of one class of states is investigated on the other class of states by using perturbation theory\cite{Lowdin}.

For low energy limit, our favorite is a Hamiltonian projected on the non-dimer sites, so it is essentially possible to divide the whole Hamiltonian into two classes namely as non-dimer classes $A_{d}-B_{u}$ and dimer classes $A_{u}-B_{d}$. An introduction of this method can be found in Ref.[\onlinecite{Rolan2003}]. In this method, Hamiltonian is defined as $H=H_0+H^{\prime}$ where $H_0$ is unperturbed Hamiltonian with the known eigenvalues $ E_n $ and eigenvectors $ |\psi_n> $ and $H^{\prime}$ is perturbed part of the Hamiltonian. The eigenvector series $\{ |\psi_n >\}$ can be departed into two weakly interacting parts called as $ \{|\psi_m>\} $ and $ \{|\psi_l >\} $ for non-dimer and dimer states, respectively.  The main idea of this method is constructing a unitary operator $e^S$ such that for the off-diagonal matrix elements of the transformed Hamiltonian ˜$\tilde{H}=e^{-s}He^S$, would be zero $< \psi_m|\tilde{H}|\psi_l>=\tilde{H}_{ml}=0$, in the non-dimer and dimer class of states. The Hamiltonian presented in Eq. \ref{Hamiltonian} can be departed into the unperturbed Hamiltonian,
\begin{equation}
\begin{aligned}
&H_0=
\begin{pmatrix}
-\dfrac{u}{2} & 0&0&0\\
0&\dfrac{u}{2}&0 &0\\
0 &0&0&\gamma_{1}\\
0&0&\gamma_{1}&0
\end{pmatrix}
\end{aligned}
\end{equation}

and the perturbed Hamiltonian,
\begin{equation}
\begin{aligned}
&H^{\prime}=
\begin{pmatrix}
0&v_{3}\pi & -v_{4}\pi^{\dagger} &v_{d}\pi^{\dagger}\\
v_{3}\pi^{\dagger}&0&v_{u}\pi& -v_{4}\pi\\
-v_{4}\pi&v_{u}\pi^{\dagger}&\dfrac{u}{2}&0\\
v_{d}\pi&-v_{4}\pi^{\dagger}&0&-\dfrac{u}{2}
\end{pmatrix}.
\end{aligned}
\end{equation}

The effective Hamiltonian projected on the non-dimer states is derived as the following successive approximations: $ \hat{H}_{eff}=H^{(0)}+H^{(1)}+H^{(2)}+H^{(3)}+... $, where the matrix elements in the non-dimer class of states is extracted from the following formulation proved in Ref.[\onlinecite{Rolan2003}].

\begin{equation}
\begin{aligned}
 & H^{(0)}_{mm^{\prime}}=(H_{0})_{mm^{\prime}},
\\
&H^{(1)}_{mm^{\prime}}=H^{\prime}_{mm^{\prime}},\\
&H^{(2)}_{mm^{\prime}}=\dfrac{1}{2}\sum_{l}H^{\prime}_{ml}H^{\prime}_{lm^{\prime}}[\dfrac{1}{E_{m}-E_{l}}+\dfrac{1}{E_{m^{\prime}}-E_{l}}],\\
&...
\end{aligned}
\label{Lowdin_formula}
\end{equation}
where the indices $ m,m^{\prime} $ scan the non-dimer class of states and $ l $ counts the dimer states. For low energies which is so close to the Dirac point, electrons momentum is too small and is considered up to quadratic terms. Therefore, the perturbative terms $H^{(j)}$ are ignored for $ j>2 $. After simple algebraic calculations, perturbative terms are derived in the second perturbative approximation. By applying the approximation $ v_{4}, v_{3}, u<\gamma_{0},\gamma_{1} $ and ignoring quadratic terms of $v_4$ and $u$ in the analytical terms, one can show that the effective Hamiltonian is exactly the same as Eq. \ref{effective_hamil}. The necessary condition for the validity of Eq. \ref{Lowdin_formula} is as the following\cite{Lowdin}, $|H^{\prime}_{ml}/(E_m-E_l)\ll 1$. This condition is satisfied by considering small values of electrons momentum and also the approximations $ v_{4}, v_{3}, u<\gamma_{0},\gamma_{1} $.

\end{document}